\documentclass[preprint,12pt, nofootinbib]{revtex4}
\usepackage[brazil, english]{babel}
\usepackage[utf8]{inputenc}
\usepackage[normalem]{ulem}
\usepackage{graphicx}
\usepackage{epsfig}
\usepackage{amssymb}
\usepackage{amsmath} 
\usepackage{slashed}
\usepackage[unicode=true,bookmarks=false,breaklinks=false,pdfborder={0 0 1},colorlinks=true]
 {hyperref}
\hypersetup{
 citecolor=blue,linkcolor=blue,urlcolor=blue}

\graphicspath{%
    {converted_graphics/}
    {/}
}
\begin{document}

\title{A new rotating axionic AdS$_4$ black hole dressed with a scalar field}

\author{Mois\'es Bravo-Gaete$^{1}$}
\email[Corresponding author: ]{moisesbravog@gmail.com, mbravo@ucm.cl} 
\author{Fabiano F. Santos$^{2,3}$}
\email[]{fabiano.ffs23@gmail.com}
\author{Jhony A. Herrera-Mendoza$^{4,5}$}
\email[]{jhonyahm@gmail.com}
\author{Daniel F. Higuita-Borja$^{5,6}$}
\email[]{dfhiguit@gmail.com}
\affiliation{
$^1$Departamento de Matem\'atica, F\'isica y Estad\'istica, Facultad de Ciencias
B\'asicas, Universidad Cat\'olica del Maule, Casilla 617, Talca, Chile.}

\affiliation{$^2$School of Physics, Damghan University, Damghan, 3671641167, Iran.}

\affiliation{$^3$ Departamento de Física, Universidade Federal do Maranhão, São Luís, 65080-805, Brazil.}

\affiliation{
$^4$Escuela de F\'isica, Facultad de Ciencias, Universidad Nacional Aut\'onoma de Honduras,
Blvr. Suyapa, Tegucigalpa, Municipio del Distrito Central 11101, Honduras.}

\affiliation{$^5$Instituto de Física, Benem\'erita Universidad Aut\'onoma de Puebla, Edificio IF-1, Ciudad Universitaria, Puebla, Pue. 72570, M\'exico,}

\affiliation{
$^6$Instituto de F\'isica, Universidad de Antioquia, Calle 70 No 52-21, Medell\'in, Colombia.}

\begin{abstract}
This paper presents a new four-dimensional {axionically charged} rotating black hole with a scalar field, which is defined by a structural function coupling the axionic field and a scalar potential.  This configuration is characterized by an integration constant and two {constant} parameters. The thermodynamic {quantities} are obtained via the Euclidean procedure, where the validity of the first law of thermodynamics is ensured. These results indicate that the rotating configuration provides a useful framework for exploring holographic superconductors, where the angular {constant} parameter plays a central role.
\end{abstract}

\maketitle

\section{Introduction}

Scalar fields non-minimally coupled to gravity {have been a central topic in black hole (BH) physics since the pioneering works} of N. Bocharova, K. Bronnikov, V. Melnikov \cite{Bocharova:1970skc} and J. Bekenstein \cite{Bekenstein:1974sf,Bekenstein:1975ts}, {which connected such configurations to the no-hair conjecture} \cite{Israel:1967wq,Israel:1967za,Carter:1971zc,Hawking:1971tu}. Although for the asymptotically flat situation the model has been subject to debate from a physical point of view, and even the existence of solutions \cite{Bronnikov:1978mx,Xanthopoulos:1992fm,Klimcik:1993cia}, the inclusion of a cosmological constant allows us to explore static BH configurations with a scalar field with other asymptotic behaviors \cite{Martinez:2002ru,Barlow:2005yd,Martinez:1996gn,Martinez:2005di,Dotti:2007cp,Charmousis:2009cm,Anabalon:2012tu,Bardoux:2012tr,Ayon-Beato:2015ada,Ayon-Beato:2024xgp,Ayon-Beato:2023bzp,Cisterna:2021xxq,Bravo-Gaete:2021hza}. For a scalar field represented by $\phi$, the Lagrangian $\mathcal{L}_{\phi}$ typically takes on the following structure:
\begin{eqnarray}\label{eq:Lphi}
\mathcal{L}_{\phi}=-\frac{1}{2}\nabla_{\mu} \phi \nabla^{\mu} \phi-\frac{1}{2}\xi R \phi^{2}-U(\phi),
\end{eqnarray}
where $\xi$ denotes the non-minimal coupling constant, $R$ is the scalar curvature, and $U(\phi)$ denotes the potential depending on $\phi$.

The construction of BHs with non-minimal scalar couplings and rotation is technically challenging, which explains the scarcity of exact solutions in the literature (see Refs. \cite{Natsuume:1999at,Correa:2012rc,Astorino:2014mda,Bardoux:2013swa,Anabalon:2009qt,Erices:2017izj,Baake:2020tgk,Cardenas:2022jtz,Barrientos:2022yoz,Bravo-Gaete:2025xya}). Nevertheless, as shown in \cite{Zhao:2013isa,Zou:2014gla}, a spinning configuration can be derived for the three-dimensional scenario where the model is represented through the action
\begin{eqnarray}\label{eq:action}
\int d^3x \sqrt{-g}\left[\frac{1}{2} (R-2\Lambda)+\mathcal{L}_{\phi}\right]=\int d^3x \sqrt{-g} (\mathcal{L}_{g}+\mathcal{L}_{\phi}),
\end{eqnarray}
with an appropriate structure for the potential $U(\phi)$ as well as $\xi=\frac{1}{8}$. Here, the equations of motion with respect to the metric $g_{\mu \nu}$ as well as the scalar field $\phi$ take the structure:
\begin{eqnarray}\label{eq:field_eqs}
\mathcal{E}_{\mu \nu}&:=& G_{\mu\nu} + \Lambda g_{\mu\nu}- T^{\phi}_{\mu\nu}=0,\\
\mathcal{E}_{\phi}&:=&  \square \phi -\xi R\phi- \frac{d U} {d \phi}=0,
\end{eqnarray}
where
\begin{eqnarray}\label{eq:Tphi}
&&  T^{\phi}_{\mu\nu} = \nabla_{\mu}\phi \nabla_{\nu} \phi -
g_{\mu\nu}\left( \dfrac{1}{2} \nabla_{\sigma} \phi \nabla^{\sigma} \phi + U(\phi) \right) + \xi \left(g_{\mu\nu}\square \phi^2 - \nabla_{\mu}\nabla_{\nu} \phi^2 + G_{\mu\nu} \phi^2\right).
\end{eqnarray}
{This motivates the search for analogous rotating configurations in four dimensions; that is,}  starting with a metric {ansatz} of the form \cite{Klemm:1997ea,Lemos:1995cm}
\begin{equation}\label{eq:metric}
ds^2=-N(r)^2f(r)dt^2+\frac{dr^2}{f(r)}+r^2(d\theta+ N^{\theta}(r)dt)^2+r^2 d \varphi^2,
\end{equation}
where $t \in \mathbb{R}$, $r$ is non-negative, while we consider the scalar field as $\phi=\phi(r)$ and set  $\xi= \frac{1}{6}$. {Here we note that the line element (\ref{eq:metric}) describes a
configuration with a planar (or toroidal) horizon geometry, and therefore the metric component $g_{\varphi\varphi}$ does not include
a factor of $\sin^{2}\theta$, as would be the case for the spherical situation.} Further, if we take into account the field equations, which are listed in Appendix \ref{EM}, we can obtain the expression for the scalar field $\phi$ from the combinations $\mathcal{E}^{t}_{\phantom{a} t}-\mathcal{E}^{r}_{\phantom{a} r}=0=\mathcal{E}^{t}_{\phantom{a} \theta}$. Nevertheless, the combination $\mathcal{E}^{\theta}_{\phantom{a} \theta}-\mathcal{E}^{\varphi}_{\phantom{a} \varphi}=0$ yields 
\begin{eqnarray}
\frac{1}{12} r^2 (\phi^2-6)\big((N^{\theta})'\big)^2=0,
\end{eqnarray}
fixing $N^{\theta}(r)=N_0$ as an integration constant and, through a coordinate transformation $\tilde{\theta}=\theta+N_0 t$, the static configuration is naturally recovered. {Here, for our notations, $(')$ denotes the derivative with respect to coordinate $r$.}

This result shows that the previous action (\ref{eq:action}) is insufficient, motivating us to search for more appropriate matter sources.
One approach is to adopt the concepts from \cite{Bardoux:2012tr,Bardoux:2012aw,Caldarelli:2013gqa}, where two 3-forms that originate from Kalb-Ramond potentials are introduced, allowing us to obtain a planar version of the hairy BH given in \cite{Martinez:2002ru,Martinez:2005di}. In addition, it has recently been shown that when the source is composed of two axionic fields that vary linearly with the coordinates of the planar base manifold, charged planar Anti-de Sitter (AdS) BHs can arise \cite{Cisterna:2018hzf,Cisterna:2019uek}, enabling the computation of the DC conductivities due to the linear dependence on the coordinates for the axionic fields, providing a momentum dissipation mechanism \cite{Andrade:2013gsa,Donos:2014cya,Donos:2013eha}, and even the potential to explore other contexts (see, for instance, Refs. \cite{Ling:2016lis,Cisterna:2017jmv,Wang:2018hwg,Bhatnagar:2017twr,BravoGaete:2019rci,Hernandez-Vera:2024zui}). The addition of a scalar field with an axionic profile has allowed the study of black strings in dynamical Chern-Simons modified gravity and their thermodynamics \cite{Corral:2021tww,Cisterna:2018jsx}, as well as in more general toy models, such as Horndeski \cite{Arratia:2020hoy}. {In this work, we show that a new four-dimensional rotating configuration arises when the action (\ref{eq:action}) is supplemented by a single axionic field $\psi$ coupled via a function $\varepsilon(\phi)$}:
\begin{equation}\label{eq:Lpsi}
\mathcal{S}_\psi[g_{\mu \nu},\psi] = -\frac{1}{2}\int d^4x \sqrt{-g}\,
{\varepsilon(\phi)} {\nabla_{\mu} \psi \nabla^{\mu} \psi},
\end{equation}
{where the term axionic refers to a shift-symmetric scalar field $\psi\rightarrow\psi+\mathrm{const.}$ that varies linearly with a spatial
coordinate. This terminology was introduced in \cite{Andrade:2013gsa,Donos:2014cya,Donos:2013eha}.}

On the other hand, over the past few decades, the study of {BHs} in AdS spacetime has gained significant interest due to its crucial role in the AdS/Conformal Field Theory (CFT) correspondence \cite{Maldacena:1997re,2840490}.
This duality unveils a fascinating relationship between {a} gravitational theory and a strongly coupled quantum field theory that lives at the conformal boundary of the bulk gravitational spacetime. This duality provides a framework to explore strongly coupled quantum systems beyond perturbative methods.
Examples of these systems are prevalent in the domains of condensed matter physics, in which high-temperature superconductors figure as prototype systems on which duality can be applied. In this regard, the seminal work in holographic superconductivity successfully modeled a high-temperature superconductor using an Abelian-Higgs-like system in a non-back-reacted planar Schwarzschild-AdS$_4$ BH background \cite{Hartnoll:2008vx}.
This groundbreaking model elucidated the mechanism of holographic superconductivity by investigating the onset of the superconducting state driven by a $U(1)$ symmetry {breaking} and analyzing the linear response of the system to electric perturbations.

Subsequent studies incorporated an external magnetic field {into} the superconducting system, enabling a holographic description of the London equation and the Abrikosov vortex lattice, which is characteristic of type-II superconductors \cite{Nakano:2008xc, Hartnoll:2008kx, Montull:2009fe, Domenech:2010nf, Maeda:2009vf, Montull:2011im, Montull:2012fy, Salvio:2012at, Donos:2020viz, delaCruz-Lopez:2024chw}. Notably, the dynamics of these holographic vortex lattices have been explored using both numerical and analytical techniques, uncovering the effect of the external magnetic field on the arrangement and density of vortices \cite{Xia:2021jzh, Herrera-Mendoza:2022whz}.  
Further realizations of holographic superconductors have considered non-commutative BH backgrounds \cite{delaCruz-Lopez:2024ync}, and the effects of non-relativistic scaling symmetries using BHs with non-standard asymptotic behaviors \cite{Bu:2012zzb,Lu:2013tza,Natsuume:2018yrg,Zhao:2013pva}. Additionally, holographic superconductivity on isotropic and anisotropic rotating backgrounds has been studied in \cite{Herrera-Mendoza:2022whz, Herrera-Mendoza:2024vfj, Sonner:2009fk, Lin:2014tza, Srivastav:2019ixc}, revealing that the BH rotation affects the properties of the superconducting state in such a way that the condensation of the scalar field is suppressed with increasing rotation. In \cite{Herrera-Mendoza:2024vfj}, a connection was suggested between the rotation of a 5-dimensional anisotropic BH and the damping effects of {quasiparticles} in a superconducting material. 

Taking all of the above into consideration, in the present {paper, we will show that (i) a} new four-dimensional rotating configuration arises upon the inclusion of an axionic field, represented by the action (\ref{eq:Lpsi}), in addition to a structured form for the potential $U(\phi)$.  {In our construction,} the scalar potential and the coupling {function} $\varepsilon(\phi)$ were not postulated a priori but reconstructed consistently to ensure integrability of the system, playing a key role in supporting the rotating configuration and exhibiting nontrivial behavior. {This reconstruction method has been performed in the literature (e.g., \cite{Erices:2017izj,Cisterna:2019uek,Cisterna:2018hzf})). The resulting BH depends only on an integration constant and two constant parameters.} Along with the previously mentioned, via the Euclidean procedure \cite{Gibbons:1976ue, Regge:1974zd}, the thermodynamic {quantities} are obtained, where the validity of the first law of thermodynamics is ensured for this spinning hairy BH.  With the above information, we will see that (ii) this rotating BH is a good laboratory for {a holographic perspective, 
where the condensation and electrical conductivity represent new behavior associated with the interaction of rotation, momentum dissipation, and non-minimal couplings, revealing effects originating from the rotating axionic background.}

Two fundamental dimensions underpin the novelty of the current work. Firstly, the inclusion of an axionic field (\ref{eq:Lpsi}) in the toy model (\ref{eq:action}):
\begin{equation}\label{eq:Ltotal}
\mathcal{S}_{\tiny{\mbox{total}}}[g_{\mu \nu}, \phi, \psi]=\int d^4x \sqrt{-g} (\mathcal{L}_{g}+\mathcal{L}_{\phi})+\mathcal{S}_{\psi},
\end{equation}
provides opportunities to obtain a rotating configuration as well as allows us to study the thermodynamic properties outside the three-dimensional scenario. {It is noteworthy  that the kinematic sector of the Lagrangian density can be written as a nonlinear $\sigma-$model \cite{Gell-Mann:1960mvl}
\begin{equation*}
 -\frac{1}{2}\nabla_{\mu} \phi \nabla^{\mu} \phi-\frac{1}{2}\varepsilon(\phi)\nabla_{\mu} \psi \nabla^{\mu} \psi\equiv-\frac{1}{2}g_{ab}\nabla_{\mu}\Sigma^a \nabla^{\mu}\Sigma^b, \hspace{3mm}\Sigma^a=(\phi,\psi),   
\end{equation*}
with the field space metric given by $ds^2=d\phi^2+\varepsilon(\phi)d\psi^2$. The associated scalar curvature $\mathcal{R}$ of this two-dimensional field space geometry reads
\begin{equation*}
    \mathcal{R}=-\dfrac{d}{d\phi}\left(\dfrac{d\varepsilon(\phi)/d\phi}{\varepsilon(\phi)}\right)-\dfrac{1}{2}\left(\dfrac{d\varepsilon(\phi)/d\phi}{\varepsilon(\phi)}\right)^2,
\end{equation*}
which, in general, is not constant. Unlike the typical situation in supergravity models, where scalar manifolds are often realized as symmetric coset spaces with constant curvature~\cite{Cremmer:1978km,Andrianopoli:1996cm}, {the geometry emerging in this case does not correspond to any standard maximally symmetric space or its non-compact counterparts}~\cite{Freedman:2012zz}. This makes the existence of such a configuration particularly intriguing, suggesting a nontrivial interplay between the scalar sector and the underlying dynamics.} Secondly, the inclusion of a rotational component offers an intriguing opportunity to explore holographic superconductors within the framework of the AdS/CFT correspondence.

The plan of the work is organized as follows: In Section \ref{ax-bh}, the new axionic rotating BH in four dimensions is presented, while the thermodynamic analysis of this configuration is performed in Section \ref{termo-rot}. In Section \ref{hol-sup}, from the BH configuration, we examine a holographic superconductor description. In particular, we will focus on the condensation and the electric conductivity. Finally, Section \ref{conclusions} is devoted to the conclusions and discussions.

\section{Four-dimensional spinning axionic black holes dressed with a scalar field}\label{ax-bh}

Consider the total action (\ref{eq:Ltotal}) in four dimensions with $\xi=\frac{1}{6}$. The equations of motion are given by
\begin{eqnarray}\label{eq:field_eqs-final}
\mathcal{G}_{\mu \nu}&:=& \mathcal{E}_{\mu\nu}- T^{\psi}_{\mu\nu}=0,\quad
\mathcal{G}_{\phi}:=  \mathcal{E}_{\phi}-\frac{d \varepsilon}{ d \phi} {(\nabla_{\mu} \psi \nabla^{\mu} \psi)}=0,\quad\mathcal{G}_{\psi}=\nabla_{\mu} \left(\varepsilon(\phi)\nabla^{\mu}\psi
\right) = 0,
\end{eqnarray}
where the corresponding contribution to the energy-momentum tensor takes the form:
\begin{eqnarray}\label{eq:Tpsi}
&& T^{\psi}_{\mu\nu} = \varepsilon(\phi)\left(
\nabla_{\mu}\psi\nabla_{\nu}\psi -
\dfrac{1}{2} g_{\mu\nu}{\nabla_{\sigma} \psi \nabla^{\sigma} \psi} \right).
\end{eqnarray}
For completeness, the explicit expressions for 
$T^{\psi}_{\mu\nu}$ are provided in Appendix \ref{EM}. Firstly, following Refs. \cite{Cisterna:2018hzf,Cisterna:2019uek,Andrade:2013gsa,Donos:2014cya,Donos:2013eha}, we consider an axionic field $\psi$ that varies linearly with respect to the coordinate $\varphi$, obtaining that $\mathcal{G}_{\psi}= \varepsilon (\partial^2_{\varphi} \psi /r^2)=0$, where $\partial^2_{\varphi} \psi:=\partial^2 \psi/\partial \varphi^2$, yielding:
\begin{equation}\label{eq:psi_field}
    {\psi(\varphi) = B \varphi}.
\end{equation}
Here, $B>0$ is an integration constant. Then, considering the combinations of the Einstein equations: 
\begin{eqnarray}\label{Gttheta}
\mathcal{G}_{t}^{\theta}&=&-\frac{r}{12}\left[r(\phi^2-6)(N^{\theta})''+2(N^{\theta})'(2\phi^2+r \phi \phi'-12)\big)\right]=0,\\
\mathcal{G}_{t}^{t}-\mathcal{G}_{r}^{r}&=&-\frac{1}{12} N^{\theta} r^2(\phi^2-6)(N^{\theta})''-\frac{1}{3} f \phi \phi''-\frac{1}{6}rN^{\theta} (r\phi \phi'+2\phi^2-12)( N^{\theta})'\nonumber\\
&+&\frac{2}{3}(\phi')^2 f=0,
\end{eqnarray}
allow us to obtain an explicit expression for the scalar field $\phi$, via the differential equation $$-\frac{f}{3}\Big(\phi \phi''-2(\phi')^2\Big)=0,$$
given by\footnote{Here we note that the most general solution is $\phi(r)=1/(ar+b)$, with $ a$ and $ b$ integration constants. Nonetheless, the remaining equations of motion demonstrate a relationship between these constants and the integration constant $B$.}
\begin{equation}\label{eq:phi_field}
    \phi(r) = \dfrac{\sqrt{6} B}{r+B}.
\end{equation}
From eq. (\ref{Gttheta}), we obtain the expression for the rotating function $N^{\theta}$, which reads
\begin{eqnarray}
{N^{\theta}(r)}&=& {\frac{\sqrt{\alpha}}{192}\left[\ln\left(\dfrac{2B+r}{r}\right) +\dfrac{2B(r+B)(2B^2+2Br-r^2)}{r^4}\right]} \nonumber\\
&=&\frac{\sqrt{\alpha}}{192}\left[\ln\left(\frac{\sqrt{6}+\phi(r)}{\sqrt{6}-\phi(r)}
\right) +\dfrac{2\sqrt{6}(4\sqrt{6}\phi(r)-\phi(r)^2-6)\phi(r)}{(\sqrt{6}-\phi(r))^4}\right],\label{eq:omega}
\end{eqnarray}
where $\alpha$ is a constant parameter, while the structural function is obtained via the combination 
$\mathcal{G}_{\theta}^{\theta}-\mathcal{G}_{\varphi}^{\varphi}=0,$ which is proportional to the algebraic equation:
$$ 12 (\phi+\sqrt{6})(\sqrt{6}-\phi)^5 \varepsilon(\phi)-\alpha\phi^4=0,$$
and taking the form:
\begin{align}
\varepsilon(\phi) &= {\dfrac{\alpha \phi^4}{12 (\sqrt{6}-\phi)^5(\sqrt{6}+\phi)}}.\label{eq:epsilon_coupling}
\end{align}
Finally, the solution for the metric (\ref{eq:metric}) with $N^{\theta}$ given in eq. (\ref{eq:omega}), through 
\begin{eqnarray}
\mathcal{G}_{t}^{t}-\mathcal{G}_{\theta}^{\theta}&=&-72 r^8 (r+B) \left(B+\frac{r}{2}\right)^2 f''-72r^7\left(B+\frac{r}{2}\right)B^2 f'\nonumber\\
&+&36 r^6 (r+2 B) (r^2+3 B r+4 B^2) f +\alpha B^6 (r+B)^5=0,
\end{eqnarray}
reads
\begin{eqnarray}\label{eq:black_factor}
N(r)&=&1, \qquad f(r)=\left(\dfrac{r}{\ell}\right)^2\left(1- \frac{{\ell^2} \eta}{\sqrt{\alpha}} N^{\theta}(r) +   \ell^2  (N^{\theta}(r))^2\right),
\end{eqnarray} where there $\eta$ is another constant parameter and $\ell$ is the AdS radius. With a negative cosmological constant,
\begin{equation}\label{eq:Lambda}
\Lambda = -\frac{3}{\ell^2},
\end{equation} 
the configuration is supported by a self-interacting potential $U(\phi)$ (this is $\mathcal{G}_{\mu}^{\nu}=0$ or $\mathcal{G}_{\phi}=0$), given by
\begin{align}
U(\phi) &={\dfrac{\phi^4}{12 \ell^2}+\dfrac{\sqrt{6} \eta  (\phi^2+4\phi+6)(\phi^2-4\phi+6)\phi}{1152 (\phi^2-6)}+\dfrac{\eta (\phi^4-36)}{2304}\ln\left(\dfrac{\sqrt{6}-\phi}{\sqrt{6}+\phi} \right)}\nonumber\\
&+{U_{\alpha}(\phi)},\label{eq:self_int_pot}
\end{align}
where
\begin{align}\label{eq:Ualpha}
U_{\alpha}(\phi)={\dfrac{\alpha (\phi^4-36) F(\phi)}{82944}\ln\left(\dfrac{\sqrt{6}-\phi}{\sqrt{6}+\phi} \right) +
\dfrac{ \alpha\left(\phi^4-6\sqrt{6} \phi^3+76\phi^2-36\sqrt{6}\phi+36 \right)(\phi^2+6)^2\phi^2}{18432 (\phi-\sqrt{6})^7(\phi+\sqrt{6})},}
\end{align}
and
\begin{align}
{F(\phi)=\dfrac{3}{16} \ln\left(\dfrac{\sqrt{6}-\phi}{\sqrt{6}+\phi} \right)+\dfrac{ 3\sqrt{6} (\phi^2+4\phi+6)(\phi^2-4\phi+6)\phi}{4(\phi^2-{6})^2(\phi^2+6)}}.\label{eq:F(phi)}
\end{align}
The coupling function $\varepsilon(\phi)$ (eq. (\ref{eq:epsilon_coupling})) and the potential $U(\phi)$ (eqs. (\ref{eq:self_int_pot}) and (\ref{eq:Ualpha})-(\ref{eq:F(phi)})) play a providential role in supporting the rotating configuration (see eqs. (\ref{eq:metric}), (\ref{eq:black_factor})-(\ref{eq:Lambda})). {It is important to emphasize that in an analogous way to the three-dimensional situation \cite{Zhao:2013isa,Zou:2014gla},  within the ans\"{a}tze (\ref{eq:metric}) under the assumptions $\phi=\phi(r)$, $\psi=\psi(\varphi)$ together with $\xi=\frac{1}{6}$, the field equations (\ref{eq:field_eqs-final})-(\ref{eq:Tpsi}) admit a novel rotating configuration described by the functions $\phi$, $\psi$, $N^{\theta}$, $U(\phi)$, $\varepsilon(\phi)$, and $f$, given by an only integration constant $B>0$ as well as the constant parameters $\alpha>0$ and $\eta>0$.}

The obtained new spinning solution exhibits the following relevant physical properties. Firstly, we note that the rotating parameter $\alpha$ takes an important role at the moment to explore this configuration. In fact, {at the limit $\alpha \rightarrow 0^{+}$ with $\eta \neq 0$, from equations (\ref{eq:omega}) and (\ref{eq:epsilon_coupling}), the rotating function $N^{\theta}$ as well as the coupling function $\varepsilon$ become zero, while the potential takes the form (\ref{eq:self_int_pot}) with $U_{\alpha}(\phi)=0$. Additionally, we note that the metric function $f$ and the scalar field $\phi$ do not vanish, allowing us to obtain a {new} static configuration where the line element is given by
\begin{equation}\label{eq:metric-not}
ds^2=-N(r)^2f(r)dt^2+\frac{dr^2}{f(r)}+r^2 (d\theta ^2+d \varphi^2),
\end{equation}
together with 
\begin{eqnarray}\label{eq:f_not}
f(r)=\left(\dfrac{r}{\ell}\right)^2\left[1- \frac{\ell^2 \eta}{192}\ln\left(\dfrac{2B+r}{r}\right)-\dfrac{\ell^2 \eta B(r+B)(2B^2+2Br-r^2)}{96 r^4} \right],
\end{eqnarray}
and $N(r)=1$. If, in addition, we now consider $\eta=0$, from eq. (\ref{eq:self_int_pot}) the potential becomes the conformal one ($U(\phi) \propto \phi^{4}$), obtaining the AdS space-time with planar base manifold
\begin{eqnarray}\label{eq:AdS}
ds^2_{\tiny{\mbox{AdS}}}=-\frac{r^2}{\ell^2}dt^2+\frac{ \ell^2 dr^2}{r^2}+r^2 (d\theta ^2+d \varphi^2),
\end{eqnarray}
and allowing us to obtain a stealth configuration
\begin{eqnarray}\label{eq:field_eqs-stealth}
&& \textcolor{black}{ G_{\mu\nu} + \Lambda g_{\mu\nu} =0= 
T^{\phi}_{\mu\nu} + T^{\psi}_{\mu\nu}}.
\end{eqnarray}
Here, $T^{\phi}_{\mu\nu}$ and $T^{\psi}_{\mu\nu}$ were given previously in eqs. (\ref{eq:Tphi}) and (\ref{eq:Tpsi}) respectively. It is important to note that eq. (\ref{eq:field_eqs-stealth}) have brought a lot of attention during recent years, allowing you to explore new and interesting solutions (see, for example, Refs. \cite{AyonBeato:2004ig,AyonBeato:2005tu,Ayon-Beato:2015qfa,Ayon-Beato:2013bsa,BravoGaete:2013djh,Hassaine:2013cma,Bernardo:2019yxp,Hassaine:2006gz,Alvarez:2016qky,Erices:2024iah}).
}

{With respect to the scalar field $\phi$, as indicated in equation (\ref{eq:phi_field}), we note that it is real, finite, and a monotonically decreasing function, where $\phi<6$ for $r>0$ and $B>0$, {whereas}
$$
\phi(r) \rightarrow 0 \quad \mbox{as} \quad r \rightarrow +\infty.\label{eq:asymp-phi}
$$
As we will see below, since the horizon is located at $r = r_h > 0$, the physically relevant 
region satisfies $0 < \phi < \sqrt{6}$, and the apparent pole of the functions $\varepsilon(\phi)$ and $U(\phi)$ at $\phi^2=6$ lies at $r=r_s= 0$, which corresponds to the curvature singularity inside the spinning BH.}  Additionally, when we perform an expansion for the coupling function $\varepsilon(\phi)$ and the potential $U(\phi)$, given in eqs.  (\ref{eq:epsilon_coupling}) and (\ref{eq:self_int_pot})-(\ref{eq:F(phi)}), we obtain:
\begin{eqnarray}
\varepsilon (\phi) &\simeq& \frac{\alpha \phi^4}{2592}+\frac{\alpha \sqrt{6} \phi^5}{3888} +O \left( \phi^6 \right),\\
U (\phi) &\simeq& {\frac {\phi^{4}}{12 \ell^{2}}}-\frac{\sqrt {6} \eta {\phi}^{5}}{3240}-{\frac {\alpha \phi^6}{46656}}+O \left( \phi^7 \right),
\end{eqnarray}
implying that these expressions vanish trivially when $r \rightarrow +\infty$. In relation to the metric function $f$ and $N^{\theta}$ as defined in eqs. (\ref{eq:omega}) and (\ref{eq:black_factor}), it is worth noting that at {infinity:}
\begin{eqnarray}
N^{\theta}(r) &\simeq& {\frac {\sqrt {\alpha}{B}^{3}}{18{r}^{3}}}+{\frac {\sqrt {
\alpha}{B}^{5}}{30{r}^{5}}}+O \left( \frac{1}{r^6} \right),\nonumber\\ 
f(r) &\simeq& {\frac{r^2}{\ell^2}-\frac{ B^3 \eta }{18 r}-\frac{B^5 \eta }{30 r^3}+O \left( \frac{1}{r^4} \right),}\label{eq:asymp}
\end{eqnarray}
showing that this solution is asymptotically AdS in four dimensions.

From the line element (\ref{eq:metric}), the scalar curvature $R$ reads
\begin{eqnarray}\label{eq:R}
	R&=&-\frac{3 N' f'}{N}-\frac{2 f N''}{N}-f''+\frac{r^2 \big((N^{\theta})'\big)^2}{2N^2}-\frac{4 f N'}{Nr}-\frac{4 f'}{r}-\frac{2f}{r^2} \nonumber\\
	  &=&{-\frac{12}{\ell^2}+\frac{12 \eta  N^{\theta}}{\sqrt{\alpha}}-12 (N^{\theta})^2+\frac{8 \eta r (N^{\theta})'}{\sqrt{\alpha}}-16r N^{\theta} (N^{\theta})'+\frac{\eta r^2 (N^{\theta})''}{\sqrt{\alpha}}-\frac{3}{2}r^2 \big((N^{\theta})'\big)^2}\nonumber\\
&-&2 r^2 N^{\theta}(N^{\theta})''.
\end{eqnarray}
Then, at the moment to evaluate (\ref{eq:R}) at $f$ and $N^{\theta}$, we get that around $r=0$:\footnote{It is worth mentioning that there exists the situation where $r=-2B<0$. However, this case is dismissed due to the constraints imposed by the range of the radial coordinate $r$.} 
\begin{eqnarray}
R \simeq -\frac{\alpha B^8}{192 r^8}-\frac{\alpha B^7}{96 r^7} +O \left( \frac{1}{r^6} \right),
\end{eqnarray}
providing a curvature singularity located at $r_s=0$. 
\begin{figure}[h!]
\centering
\begin{tabular}{cc}
\includegraphics[width=0.5\textwidth]{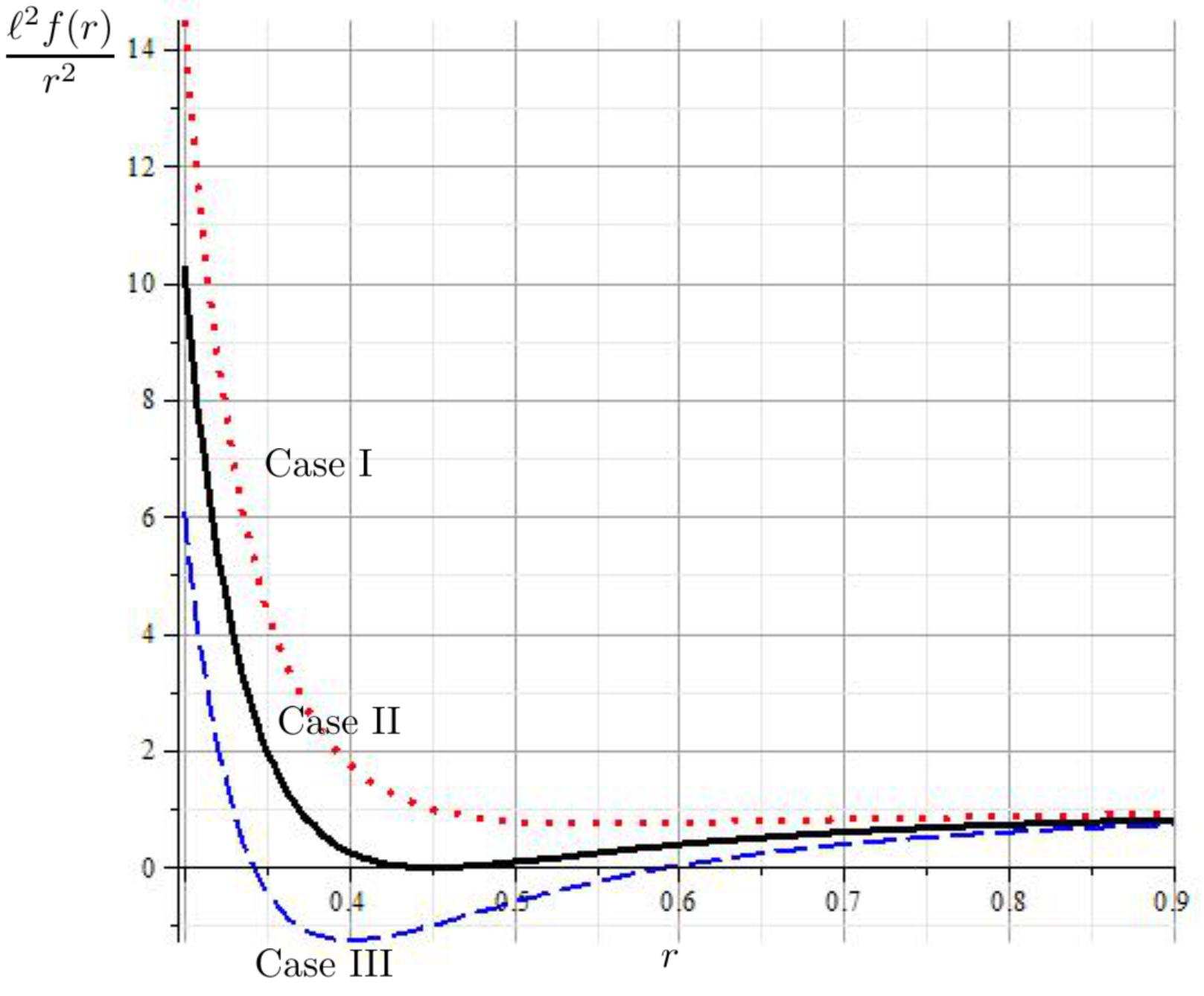} &  \includegraphics[width=0.46\textwidth]{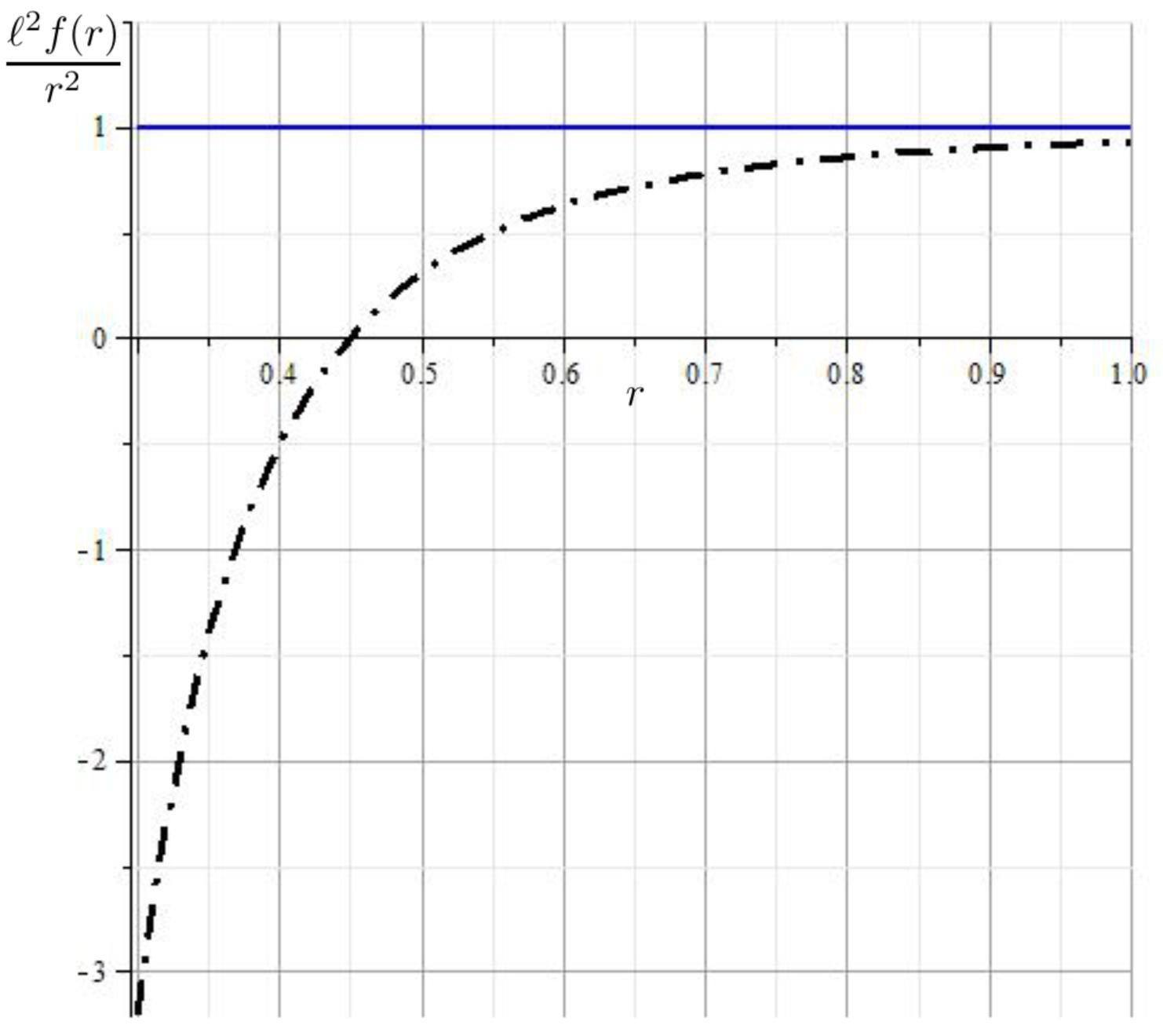}
\end{tabular}
\caption{{ Left Panel:} Graphical representation of {${\ell^2 f(r)}/{r^{2}}$} as a function of the radial coordinate $r$, with  {$\eta=B=\ell=1$} for our calculations. Here, Case I (red dotted curve), characterized by $\alpha=1$, indicates the presence of a naked singularity ($4-{\ell^2 \eta^2/\alpha}>0$).  Case II (black curve) corresponds to the extremal case, given by {$\alpha=1/4$}. Here, {${\ell^2 f(r^{*})}/{(r^{*})^{2}}=({\ell^2 f(r^{*})}/{(r^{*})^{2}})'=0$ and this condition emerges when $4-{\ell^2 \eta^2/\alpha}=0$}. Finally, Case III (dashed blue curve and when {$\alpha=1/9$}) reveals the situation where {$4-{\ell^2 \eta^2/\alpha}<0$}. {Right Panel:} Graphical representation of the {increasing function ${\ell^2 f(r)}/{r^{2}}$ at the limit $\alpha \rightarrow 0^{+}$, with  $\eta=B=\ell=1$, through the black dotted dashed curve. Here, the location of the event horizon is ensured. }}
\label{fig1}
\end{figure}

On the other hand, to study the existence of {the event horizon (or outer)}, we consider the function
$${\frac{l^2 f(r)}{r^{2}}=1- \frac{{\ell^2}\eta}{\sqrt{\alpha}} N^{\theta}(r) +  \ell^2  (N^{\theta}(r))^2},$$
where 
$$\displaystyle{\lim_{r \rightarrow 0+}\frac{\ell^2 f(r)}{r^{2}}=+\infty, \qquad \lim_{r \rightarrow +\infty}\frac{\ell^2 f(r)}{r^{2}}=1,}$$ whereas
\begin{eqnarray}\label{eq:fp}
{\left(\frac{\ell^2 f(r)}{r^{2}}\right)'=(N^{\theta}(r))' \ell^2 \left(2 N^{\theta}(r)-\frac{\eta}{\sqrt{\alpha}}\right)=-\left(\frac{\sqrt{\alpha} B^3 \ell^2 (r+B)^2}{6r^5 (2B+r)}\right) \left(2 N^{\theta}(r)-\frac{\eta}{\sqrt{\alpha}}\right)}.
\end{eqnarray}
Here, the increasing or decreasing behavior depends on whether the sign of { $2 N^{\theta}-\eta/\sqrt{\alpha}$} is negative
or positive. For the special case $r=r^{*}$ such that ${N^{\theta}(r^{*})=\eta/(2 \sqrt{\alpha})}$ , the equation (\ref{eq:fp}) vanishes, and if 
$${\displaystyle{\left(\frac{\ell^2 f(r)}{r^{2}}\right)\Big{|}_{r=r^{*}}=1-\frac{\ell^2 \eta^2}{4 \alpha} \leq 0,}}$$
the existence of horizons is ensured, as is shown in Figure \ref{fig1} left panel. For the sake of completeness, at the limit $\alpha \rightarrow 0^{+}$, considering now the metric function (\ref{eq:f_not}) and 
$$\frac{\ell^{2} f(r)}{r^{2}}=1- \frac{\ell^2 \eta}{192}\ln\left(\dfrac{2B+r}{r}\right)-\dfrac{\ell^2 \eta B(r+B)(2B^2+2Br-r^2)}{96 r^4},$$
we note that
$$\displaystyle{\lim_{r \rightarrow 0+}\frac{\ell^2 f(r)}{r^{2}}=-\infty, \qquad \lim_{r \rightarrow +\infty}\frac{\ell^2 f(r)}{r^{2}}=1},$$
whereas
$$\displaystyle{\left(\frac{\ell^2 f(r)}{r^{2}}\right)'=\frac{\eta \ell^2 B^3 (r^2+2Br+B^2)}{6 r ^5 (2B+r)}>0},$$
ensures the existence of {an} event horizon, as shown in Figure \ref{fig1} right panel.

Building upon the analysis and framework established for this new four-dimensional axionic spinning BH configuration and the presence of the integration constant $B$ as well as the {parameter} constants $\alpha$ and $\eta$, in the next section, {we} will delve into the thermodynamic quantities.

\section{Thermodynamics of the rotating black hole solution}\label{termo-rot}

The thermodynamic analysis of the previously presented rotating solution (\ref{eq:metric})-(\ref{eq:F(phi)}) will be conducted using the Euclidean procedure. In this approach, the partition function is equated with the Euclidean path integral, employing the saddle point approximation around the Euclidean continuation of the solution \cite{Gibbons:1976ue, Regge:1974zd}. In this Euclidean framework, the Euclidean time $\tau$ is introduced as an imaginary counterpart to the time coordinate $t$, linked through the relation $\tau=it$ from the metric (\ref{eq:metric}). The radial coordinate $r$ spans from the location of, for our case, the outer horizon denoted as $r_h$, to $+\infty$, while the Euclidean time is made periodic to avoid conical singularities, adhering to the range $0\leq \tau<\beta$, where $\beta=T^{-1}$ and $T$ is the Hawking temperature. {Following this standard minisuperspace procedure, once the geometric symmetries are imposed, the ansatz defines a consistent truncation of the full configuration space. Therefore, the functions are treated as independent fields of a one-dimensional effective action, and the reduced Euclidean action reads:} 
\begin{eqnarray}
I_{E}&=&  \beta \int_{r_h}^{+\infty} dr \int_{0}^{\sigma_{\theta}} d\theta \int_{0}^{\sigma_{\varphi}} d \varphi \Big\{N \Big[ \left(1-\frac{\phi^2}{6}\right) (r f'+f)-\frac{1}{6} r \phi ' (-r f \phi'+r \phi f'+4 \phi f)\nonumber\\
&-&\frac{1}{3}\phi r^2 f \phi''+\Lambda r^2+U(\phi) r^2 -\frac{6 p^2}{r^4 (\phi^2-6)}\Big]+N^{\theta} p'\Big\}\nonumber\\
&+&\frac{1}{2}\beta \int_{r_h}^{+\infty} dr \int_{0}^{\sigma_{\theta}} d\theta \int_{0}^{\sigma_{\varphi}} d \varphi \big({N} \varepsilon(\phi) \left(\partial_{\varphi} \psi\right)^{2}\big) +B_{E}, \label{euclideanaction}
\end{eqnarray}
where $B_{E}$ is a boundary term constructed such that the Euclidean
action has an extremum $\delta I_{E}=0$ and
\begin{equation}
p(r)=-\frac{r^4 \left(\phi(r)^2-6\right) N^{\theta}(r)' }{12 N(r)}. \label{p}
\end{equation}
The variation with respect to the dynamical fields $N$, $f$, $\phi$, $p$, $N^{\theta}$ and $\psi$ yield the following equations
\begin{eqnarray*}
E_{N}&:=& \left(1-\frac{\phi^2}{6}\right) (r f'+f)-\frac{1}{6} r \phi ' (-r f \phi'+r \phi f'+4 \phi f) -\frac{1}{3}\phi r^2 f \phi''+\Lambda r^2+U(\phi) r^2\nonumber\\
&+&\frac{1}{2}\varepsilon(\phi) \left(\partial_{\varphi} \psi\right)^{2} -\frac{6 p^2}{r^4 (\phi^2-6)}=0,\nonumber\\
E_{f}&:=&{\frac{1}{6} r\left(2 r (\phi')^2-r \phi  \phi''\right)N+\frac{1}{6} r\left( r \phi \phi'+\phi^2-6\right)N'=0}, \\ \\
E_{\phi}&:=&{-\frac{1}{3} \phi r^2 f N''-\frac{1}{6} r \left(6 r f \phi'+4 \phi f +3 \phi r f' \right) N'-\frac{1}{6} \Big(4 \phi r f'+2 \phi f+r^2 \phi f''+6 \phi'  r^2 f'}\nonumber\\
&+&{12 r f \phi'+6 r^2 f \phi'' -6 r^{2} U_{\phi} -3\varepsilon_{\phi} \left(\partial_{\varphi} \psi\right)^{2}+\frac{12 \phi p^2 }{r^4 (\phi^2-6)^2}  \Big) N =0},\\ \\
E_{p}&:=&12 N p+r^4(\phi^2-6) (N^{\theta})' =0,\\ \\ 
E_{N^{\theta}}&:=&p'=0, \\ \\
E_{\varphi}&:=&\partial_{\varphi}^{2} \psi=0.
\end{eqnarray*}
In our notations, $U_{\phi}={dU}/{d\phi}, \, \varepsilon_{\phi}={d\varepsilon}/{d\phi}, \, \partial_{\varphi}^{2} \psi =\partial^{2} \psi/ \partial \varphi^{2}$. Note that \textcolor{black}{$E_{N^{\theta}}=0$} implies $p=\mbox{constant}$, which is related to the angular momentum $\mathcal{J}$ as we will see in the following lines. One can check that the system of equations is consistent with the rotating {solution presented in Section \ref{ax-bh}}. From eq. (\ref{euclideanaction}), the variation of the boundary term $\delta B_{E}$ reads
\begin{eqnarray}\label{eq:deltaBe}
\delta B_{E}&=& \beta \sigma \Big[N\left(\frac{1}{6} r^2 \phi \phi'+r \left(\frac{1}{6} \phi^2-1\right)\right) \delta f-\frac{1}{6}r^2\left( N\left(4 f \phi'+\phi f'\right)+2N'\phi f\right)\delta \phi \nonumber\\
&+&\frac{N}{3} \phi  r^2 f  \delta (\phi')-N^{\theta} dp\Big]_{r_h}^{+\infty}-{\beta \int_{r_h}^{+\infty} dr N \varepsilon   \left[\int_{0}^{\sigma_{\theta}} d\theta  \partial_{\varphi} \psi \delta \psi \right]_{0}^{\sigma_{\varphi}}},   
\end{eqnarray}
where the coordinates $\theta$ and $\varphi$ are assumed in the intervals $0 \leq \theta \leq \sigma_{\theta}$ and $0 \leq \varphi \leq \sigma_{\varphi}$, while $\sigma$ represents
$$\sigma=\int_{0}^{\sigma_{\theta}} d\theta \int_{0}^{\sigma_{\varphi}} d \varphi =\sigma_{\theta} \sigma_{\varphi}.$$
Here, we emphasize that the Euclidean action used in our thermodynamic analysis is not based solely on the Ricci scalar $R$, but rather includes the full contributions from the non-minimally coupled scalar field and the axionic sector. Additionally, the Euclidean action (\ref{euclideanaction}) is constructed by evaluating the complete Lagrangian and including a boundary term $B_E$ whose variation $\delta B_E$ is derived explicitly to ensure that $\delta I_E = 0$ on-shell (see eq. (\ref{eq:deltaBe})). This boundary analysis is crucial and ensures that the thermodynamic quantities obtained via the saddle-point approximation are physically meaningful and consistent with the first law.

Now we are in a position to compute the boundary term $B_{E}$. {For this, we calculate the variation of the dynamical fields $\delta N^{\theta}$, $\delta f$, $\delta \psi$, $\delta \phi$, and $\delta p$, where the constant parameters $\alpha$ and $\eta$ are fixed. The variations are performed with respect to the integration constant $B$ as well as the location of the event horizon $r_h$. At infinity,}  we have that
\begin{eqnarray}
\delta N^{\theta} \big{|}_{+\infty} &=& {{\frac {\sqrt {\alpha}{B}^{2} \delta B}{6{r}^{3}}}+{\frac {\sqrt {
\alpha}{B}^{4} \delta B}{6{r}^{5}}}+O \left( \frac{1}{r^6} \right)},\nonumber\\ 
\delta f \big{|}_{+\infty} &=&{{ -\frac{ \eta B^2  \delta B}{6 r}-\frac{ \eta B^4  \delta B}{6 r^3}+O \left( \frac{1}{r^4} \right),
}}
\end{eqnarray}
whereas with respect to $p$, and the scalar fields ($\phi,\psi$):
\begin{eqnarray}\label{eq:varsf}
\delta\phi \big{|}_{+\infty} &=& {\frac{\sqrt{6} \delta B}{r}-\frac{2 \sqrt{6} B \delta B}{r^2}+O \left( \frac{1}{r^3} \right)},\nonumber\\ 
\delta \psi \big{|}_{+\infty} &=& \varphi \delta B, \qquad \delta p \big{|}_{+\infty}  = {-\frac{ \sqrt{\alpha} B^2 \delta B}{4}=\delta \left(-\frac{1}{12} {\sqrt{\alpha}B^3}\right).}
\end{eqnarray}
With all these ingredients, we obtain that at the infinite
$${\delta B_{E}\big{|}_{+\infty}={\frac{\beta \sigma  \eta B^{2} \delta B}{6}=\delta\left(\frac{\beta \sigma \eta B^3}{18}\right)}\Rightarrow B_{E}\big{|}_{+\infty}=\frac{\beta \sigma \eta B^3}{18}.}$$

On the other hand, considering at the horizon $r_h$ 
\begin{eqnarray}\label{eq:hor}
\delta\phi \big{|}_{r_h} &=& \delta \phi(r_h)-\phi'\big{|}_{r_h} \delta r_h, \qquad \delta \psi \big{|}_{r_h} = \varphi \delta B,\nonumber\\ 
\delta f \big{|}_{r_h} &=& -\frac{4 \pi}{\beta} \delta r_h , \qquad \delta p \big{|}_{r_h}  = \delta \left(-\frac{1}{12} {\sqrt{\alpha}B^3}\right) ,
\end{eqnarray}
we have {that the last expression of eq. (\ref{eq:deltaBe}) (with $N=1$) takes the form:
$$\int_{r_h}^{+\infty} dr \varepsilon   \left[\int_{0}^{\sigma_{\theta}} d\theta  \partial_{\varphi} \psi \delta \psi \right]_{0}^{\sigma_{\varphi}}=\int_{r_h}^{+\infty} dr (B \varepsilon) (\sigma \delta B)=\Psi_a \delta (\sigma B)  $$
and
$$\delta B_{E}\big{|}_{r_h}= \delta \left(\frac{2 \sigma \pi r_h^3 (2B+r_h)}{(r_h+B)^2}\right)+\beta \Omega \delta (\sigma p)+\beta \Psi_a \delta (\sigma B) \Rightarrow$$ $$B_{E}\big{|}_{r_h}=  \left(\frac{2 \sigma \pi r_h^3 (2B+r_h)}{(r_h+B)^2}\right)+\beta \Omega (\sigma p)+\beta \Psi_a  (\sigma B),$$
where}
$$p=-\frac{1}{12} {\sqrt{\alpha}B^3},$$
while $\Omega$ and {$\Psi_{a}$} are the chemical potentials corresponding to the angular momentum ${\cal{J}}$ and the axionic charge {${\omega}_{a}$}, identified as
\begin{eqnarray}
\Omega&=&\lim _{r \rightarrow \infty}N^{\theta}(r)-N^{\theta}(r_h)=-N^{\theta}(r_h),\\
{\Psi_{a}}&=&\int_{r_h}^{+\infty} {dr B \varepsilon\big(\phi(r)\big) =\left[-\frac{B^3 \alpha (r+B) (-r^2+2Br+2B^2)}{1152 r^{4}}- \frac{B^2 \alpha}{2304} \ln\left(\frac{2B+r}{r} \right)\right]_{r_h}^{+\infty}}\nonumber\\
&=&\frac{\sqrt{\alpha}B^2}{12} N^{\theta}(r_h).
\end{eqnarray}
With all the above, the boundary term $B_{E}$ is found to be
\begin{eqnarray}
B_{E}&=&{B_{E}\big{|}_{+\infty}-B_{E}\big{|}_{r_h}=\frac{\beta \sigma \eta B^3}{18}-\left(\frac{2 \sigma \pi r_h^3 (2B+r_h)}{(r_h+B)^2}\right)-\beta \Omega (\sigma p)-\beta \Psi_a  (\sigma B).} \label{Bound}
\end{eqnarray}
The identification of the thermodynamic parameters is
achieved {by} recalling that the Euclidean action is related to the Gibbs
free energy $G$ as
\begin{equation}\label{free_ene}
{I_{E}=\beta G = \beta{\cal{M}}-{\cal{S}}-\beta \Omega {\cal{J}}-\beta \Psi_{a} \omega_{a}},
\end{equation}
where ${\cal{M}}$ is the mass, ${\cal{S}}$ the entropy, and, as before,  $\Omega$ (respectively {$\Psi_a$)} is the chemical potential associated with the angular momentum ${\cal{J}}$ (respectively axionic charge  {$\mathcal{\omega}_a$}).
Hence, comparing (\ref{Bound}) with (\ref{free_ene})
one obtains that the entropy reads
\begin{eqnarray}\label{entropy}
{\cal{S}}=2\pi \sigma  r_h^2 \left(1-\frac{1}{6}\phi(r_h)^2\right)=\frac{2 \sigma \pi r_h^3 (2B+r_h)}{(r_h+B)^2},
\end{eqnarray}
while the mass ${\cal{M}}$, angular momentum ${\cal{J}}$, angular velocity $\Omega$, axionic potential {$\Psi_a$}, and axionic charge {${\cal{\omega}}_a$} are given {by
\begin{eqnarray}
{\cal{M}}&=&{\frac{\sigma  \eta B^3}{18}},\label{mass}\\
{\cal{J}}&=&{-\frac{\sigma}{12} {\sqrt{\alpha}B^3}},\qquad  \Omega=-N^{\theta}(r_h),\label{momentum-vel}\\
{\Psi_a}&=& \frac{B^2 \alpha (r_h+B) (-r_h^2+2Br_h+2B^2)}{1152 r_h^{4}}+ \frac{B \alpha}{2304} \ln\left(\frac{2B+r_h}{r_h} \right)=\frac{\sqrt{\alpha}B^2}{12} N^{\theta}(r_h),\\
{\omega_a}&=&\sigma B.
\end{eqnarray}
The Hawking temperature reads
\begin{eqnarray}\label{hawkingtemp}
T=\frac{\kappa}{2 \pi}=\frac{N(r) f'(r)}{4 \pi} \Big{|}_{r=r_{h}}={\frac{(N^{\theta}(r))'(2 \sqrt{\alpha}N^{\theta}(r) -\eta) r^2}{4 \pi \sqrt{\alpha}}\Big{|}_{r=r_{h}},}
\end{eqnarray}
where $\kappa$ is the surface gravity, given by
\begin{equation}\label{kappa}
\kappa=\sqrt{-\frac{1}{2}\left(\nabla_{\mu} \xi_{\nu}\right)\left(\nabla^{\mu} \xi^{\nu}\right)},
\end{equation}
with a Killing vector $\xi^{\mu}=(\partial_t)^{\mu}-\Omega (\partial_\theta)^{\mu}$ and $\Omega=-N^{\theta}(r_h)$. Here, it is important to note that:
\begin{eqnarray*}
\delta{\cal{M}}&=&\frac{ \sigma  \eta B^{2} \delta B}{6},\nonumber\\
T \delta{\cal{S}}&=& T (\partial_{r_h}{\cal{S}}) \delta r_h+ T (\partial_{B}{\cal{S}}) \delta B=\frac{ \sigma  \eta B^{2} \delta B}{6}-{\frac{\sigma N^{\theta}(r_h)}{3} {\sqrt{\alpha}B^2}} \delta B, \nonumber\\
\Omega \delta {\cal{J}}&=& {\frac{\sigma N^{\theta}(r_h)}{4} {\sqrt{\alpha}B^2}} \delta B, \nonumber\\
\Psi_a \delta {\omega_a}&=& {\frac{\sigma N^{\theta}(r_h)}{12} {\sqrt{\alpha}B^2}} \delta B,
\end{eqnarray*}
ensuring the first law
\begin{eqnarray}\label{eq:first-law}
{\delta {\cal{M}}=T \delta {\cal{S}}+\Omega \delta {\cal{J}}+\Psi_a \delta {\omega_a}}.
\end{eqnarray}
Here, we are using the fact that from eqs. (\ref{eq:omega}) and (\ref{eq:black_factor}): 
$$ 1- \frac{{\ell^2}\eta}{\sqrt{\alpha}} N^{\theta}(r_h,B) +  \ell^2  (N^{\theta}(r_h,B))^2=0,$$
and from this last expression:
$$ \left(\partial_{r_h} N^{\theta}\right) \delta r_h+\left(\partial_{B} N^{\theta}\right) \delta B=0,$$
with $\partial_{a} {\cal{S}}:=\partial {\cal{S}}/ \partial a $ and $\partial_{a} N^{\theta}:=\partial N^{\theta} / \partial a$. To reinforce the above, the entropy ${\cal{S}}$ (eq. (\ref{entropy})) can be recovered using the Wald formula \cite{Wald:1993nt,Iyer:1994ys}, denoted as ${\cal{S}}_{W}$, where
\begin{eqnarray*}
\label{wald} {\cal{S}}_{W}=-2 \pi \oint_{\Sigma} d ^2x \sqrt{|\gamma|} P^{\mu \nu \sigma \rho} \, \varepsilon_{ \mu \nu} \,  \varepsilon_{\sigma \rho}=2\pi \sigma  r_h^2 \left(1-\frac{1}{6}\phi(r_h)^2\right)=\frac{2 \sigma \pi r_h^3 (2B+r_h)}{(r_h+B)^2},
\end{eqnarray*}
and
\begin{equation*}\label{P}
P^{\mu \nu \sigma \rho }=\frac{\partial (\mathcal{L}_{g}+\mathcal{L}_{\phi})} {\partial R_{\mu \nu \sigma \rho}}=\frac{1}{4} \left(1-\frac{1}{6}\phi^2\right)\left(g^{\mu \sigma} g^{\nu \rho}-g^{\mu \rho} g^{\nu \sigma}\right).   
\end{equation*}
Here, the expression is evaluated at the spatial section $\Sigma$ of the event horizon. The quantity $|\gamma|$ represents the determinant of the induced metric on $\Sigma$, while $\varepsilon_{\mu \nu}$ denotes the binormal vector
$$\varepsilon_{\mu \nu}=-\varepsilon_{\nu \mu}:=\frac{1}{\kappa}\,\nabla_{\mu} \xi_{\nu},$$
where  $\kappa$ is the surface gravity shown previously in eq. (\ref{kappa}).} The mass $\mathcal{M}$ and angular momentum $\mathcal{J}$  can be recovered via the Brown-York method \cite{Brown:1992br,Brown:1994gs,Balasubramanian:1999re}, where the explicit steps are given in Appendix \ref{termo-b}.

For the non-spinning situation, represented via the line element (\ref{eq:metric-not}) with  $N=1$, together with the metric function $f$ (\ref{eq:f_not}), the scalar field $\phi$ (\ref{eq:phi_field}) and the potential $U(\phi)$  (\ref{eq:self_int_pot}) with $U_{\alpha}(\phi)=0$, we note that  the entropy ${\cal{S}}$ as well as  the mass ${\cal{M}}$ take the form given in (\ref{entropy}) and (\ref{mass}) respectively, while {the Hawking temperature $T$} reads 
$$T=\frac{\eta B^3 (r_h+B)^2}{24 \pi r_h^3 (2 B+r_h)},$$
and the first law $\delta {\cal{M}}=T \delta {\cal{S}}$ holds, where now $r_h$ corresponds to the location of the event horizon for the non-spinning configuration.

\section{Holographic superconductor}\label{hol-sup}

In this section, we are interested in exploring a holographic superconductor description of the preceding {novel rotating axionic AdS$_4$ BH, investigating how holographic superconductivity behaves in the presence of both rotation and momentum dissipation. To our knowledge, this particular combination of rotation, momentum dissipation, and non-minimal scalar coupling has not been explored before.} Given this, we find it convenient to focus on a particular case of this background in which the transformation $u\to r_h/r$ makes it feasible to locate the outer horizon at $u=1$ and the asymptotic boundary at $u=0$. This is possible, provided we define
\begin{equation}\label{eq:conditions}
	{\eta = \left(\dfrac{384}{\ln 3}\right)\dfrac{1}{\ell^2}+\left(\dfrac{\ln 3}{384}\right)\alpha}, \qquad B=(\sqrt{3}-1) r_{h},
\end{equation}
which preserves the genuine nature of the {constant} parameter $\alpha$ as a rotation parameter. In this manner, by evaluating the above expressions into the background functions \eqref{eq:black_factor}-\eqref{eq:F(phi)} and taking the limit $\alpha\to 0^{+}$, the angular momentum of the background is turned off and a static AdS$_{4}$ BH configuration is obtained (see eqs. (\ref{eq:metric-not})-(\ref{eq:f_not})). Moreover, in the setting \eqref{eq:conditions}, the Hawking temperature \eqref{hawkingtemp} adopts the simpler form
\begin{equation}\label{eq:Hawking_temp2}
	T= \dfrac{(3-\sqrt{3})(147456-\ell^2 (\sqrt{\alpha} \ln 3)^2 ) r_h}{221184\, \pi\, (\ln 3)\, \ell^2 }.
\end{equation}
With this in mind, to construct the superconductor, we consider an additional matter contribution consisting of a Maxwell-complex scalar field system of the form \cite{Gubser:2008px, Hartnoll:2008vx,Hartnoll:2008kx}
\begin{equation}\label{eq:matter_sup}
	S_{\text{M}}=-\frac{1}{q^2}\int{d^{4}x \sqrt{-g}\Biggl(\dfrac{1}{4}F_{\mu\nu}F^{\mu\nu}+\dfrac{1}{\ell^2}\left(|D_\mu\Psi|^2+m^2|\Psi|^2\right)\Biggr)},
\end{equation}
in which $F_{\mu\nu} = 2\partial_{[\mu}A_{\nu]} $, while $\Psi$ denotes a complex scalar field characterized by mass $m$ and charge $q$, minimally coupled to the gauge field $A_{\mu}$ through the extended covariant derivative $D_{\mu} = \nabla_{\mu} -i A_{\mu}$. Here, we work in the probe limit $q\gg1$, where the backreaction of the matter contribution \eqref{eq:matter_sup} onto the background fields can be effectively ignored.  
Thus, the {coupled} system of field equations is given by 
\begin{subequations}\label{eq:GenFieldEqns_sup}
	\begin{align}
		\nabla_\mu F^{\mu\nu}-\dfrac{1}{\ell^2}\left(2A^\nu|\Psi|^2-i\Psi\nabla^\nu\overline{\Psi}+i\overline{\Psi}\nabla^\nu\Psi\right) & =0,\label{eq:gen_max_sup}\\
		\nabla^{\mu}\nabla_{\mu}\Psi-2iA^{\mu}\nabla_{\mu}\Psi-i\Psi\nabla_{\mu}A^{\mu}-\left(m^2+A_{\mu}A^{\mu}\right)\Psi & =0,\label{eq:gen_sca_sup}
	\end{align}
\end{subequations}
which is intended to be solved on the fixed background \eqref{eq:metric}, taking into account eq. \eqref{eq:conditions}. {This coupling system is a direct consequence of the structure of the background solution, where the scalar and axionic sectors interact with the metric. Once the background is fixed, the system is solved simultaneously with the standard boundary conditions at the horizon and at the AdS boundary.}

\subsection{Condensation of the scalar field and the electric conductivity}

In the Abelian Higgs-like description of holographic superconductors as outlined in eq. \eqref{eq:matter_sup}, the mechanism responsible for the onset of superconductivity is the spontaneous breaking of the $U(1)$ Abelian gauge symmetry. This symmetry-breaking paves the way for a non-zero profile of the scalar field. Consequently, in this context, the scalar field is identified as the order parameter that undergoes condensation below some critical temperature.
Here, we would like to investigate whether the scalar field $\Psi$ possesses such a property. To do so, using the gauge symmetry of  \eqref{eq:matter_sup}, it is sufficient to consider a real scalar field and a Maxwell field with the simplest dependence 
\begin{equation}\label{eq:ansatz_condensate}
	\Psi=\Psi(u), \qquad {A=A_t(u) dt + A_{\theta}(u) d\theta}.
\end{equation}
In this setting, the system of equations \eqref{eq:gen_max_sup}-\eqref{eq:gen_sca_sup} reduces to
\begin{subequations}\label{eq:CondeFieldEqns}
	\begin{align}
		&A_t''+\ell^2 \dfrac{N^{\theta} ({N^{\theta}})'}{f}A_t'+\left[\dfrac{N^{\theta} f'}{f}-\dfrac{({N^{\theta}})'}{f}\left(\ell^2 ({N^{\theta}})^2+f\right)\right]A_{\theta}'-\left(\dfrac{2\psi^2}{u^2f}\right)A_t  =0,\\
		&A_{\theta}''+\left(\dfrac{f'}{f}-\ell^2\dfrac{N^{\theta} ({N^{\theta}})'}{f}\right)A_{\theta}'+\ell^2 \dfrac{({N^{\theta}})'}{f}A_t'-\left(\dfrac{2\psi^2}{u^2f}\right)A_{\theta}  =0,\\
		&\Psi''+\left(\dfrac{f'}{f}-\dfrac{2}{u}\right)\Psi'-\dfrac{\ell^2}{u^2r_h^2f}\Biggl(m^2 r_h^2+u^2 A_\theta^2-\ell^2 u^2\dfrac{\left(A_t-N^{\theta} A_\theta\right)^2}{f}\Biggr)\Psi  =0,
	\end{align}
\end{subequations}
where now the primes $(')$ stand for derivatives with respect to $u$. {Eqs. (\ref{eq:CondeFieldEqns}a)-(\ref{eq:CondeFieldEqns}c) remain coupled because the function $N^{\theta}$ mixes temporal and spatial components of the gauge field $A_\mu$. This is a characteristic feature of rotating holographic backgrounds, and it is precisely this mixing that allows the parameter $\alpha$ to influence the superconducting phase.} In solving this system of equations, we need to specify appropriate boundary conditions at the horizon ($u=1$) and at the conformal boundary of the background ($u=0$). Concretely, at the horizon, we require that the scalar field is finite $\Psi(1)<\infty$, while the components $A_{t}(1)$ and $A_{\theta}(1)$ vanish for $A_{\mu}$ to have a finite norm everywhere. On the other hand, at the boundary, the fields behave as 
\begin{equation}\label{eq:BoundarySols}
	\Psi=\Psi_{1}u^{\Delta_{1}}+\Psi_{2}u^{\Delta_{2}}, \quad A_t=\mu-\rho\left(\dfrac{u}{r_h}\right), \quad A_{\theta}=\mu_{\theta}-\mathcal{J}_{\theta} \left(\dfrac{u}{r_h}\right), 
\end{equation}
with
\begin{equation}\label{eq:scaling_dim}
	\Delta_{1,2}= \dfrac{1}{2}\left(3\pm \sqrt{9+4m^2\ell^2}\right).
\end{equation}
From the holographic standpoint, the coefficient $\Psi_{1}$ (respectively $\Psi_{2}$) is identified as the order parameter $\mathcal{O}$ with scaling dimension $\Delta_{1}$ (respectively $\Delta_{2}$). However, to have well-defined normalizable modes at the Ultraviolet (UV) regime (boundary), we must restrict the scalar field mass to be bounded below as $m^2\ell^2 \ge -\frac{9}{4}$, which defines the Breitenlohner-Freedman (BF) bound for a massive scalar field in AdS$_4$  \cite{Breitenlohner:1982bm,Breitenlohner:1982jf}. Besides, $\mu$ and $\rho$ are respectively interpreted as the chemical potential and the charge density operator. A similar interpretation is given to the coefficients of $A_{\theta}$, where $\mu_{\theta}$ stands for a one-dimensional potential vector, {the source} of the density of {the current} operator $\mathcal{J_{\theta}}$.

Using the holographic dictionary and by numerically solving the system \eqref{eq:CondeFieldEqns}, we determine the condensation profile of the order parameter as a function of the temperature.
{ For our analysis, we work in the canonical ensemble, keeping the charge density $\rho$ fixed. Moreover, we set the scalar mass to $m^2\ell^2 = -2$, a value consistent with the BF bound, and that ensures the simplest asymptotic structure for the scalar field. Concretely, this choice sets the scaling dimensions of the dual operators to $\Delta_{1} = 1$ and $\Delta_{2} = 2$ for $\mathcal{O}_{1}$ and $ \mathcal{O}_{2}$, respectively. 

\begin{figure}[h!]
	\centering
\includegraphics[width=.49\textwidth]{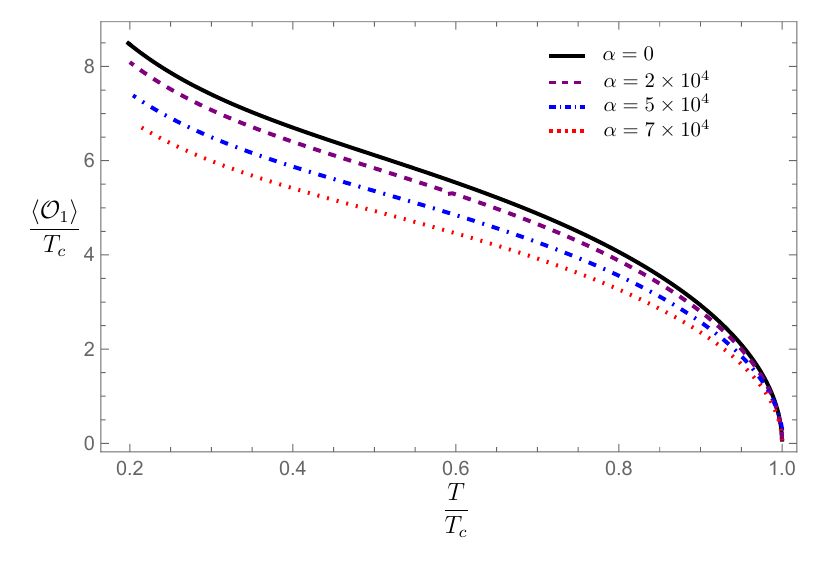}
\includegraphics[width=.5\textwidth]{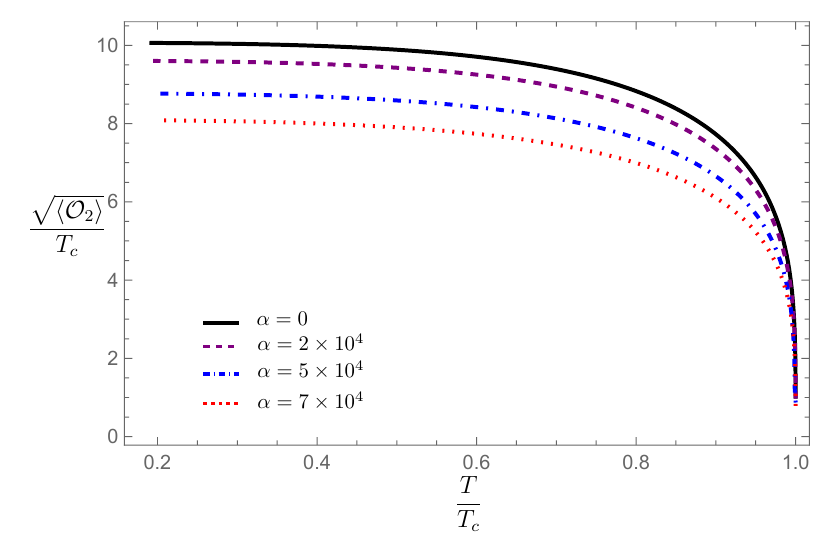}
	\caption{The condensation profiles for the operators $\mathcal{O}_1$ and $\mathcal{O}_2$ as a function of temperature, considering different values of the rotation parameter $\alpha$. The continuous (black) curve corresponds to the static case, while the dashed (violet), dot-dashed (blue), and dotted (red) {curves} correspond to situations with non-zero rotation.}
	\label{fig:Condensate}
\end{figure}

In Fig. \ref{fig:Condensate}, we illustrate the influence of the background rotation on the condensation curves for  $\mathcal{O}_1$ (left panel) and $\mathcal{O}_2$ (right panel). We observe that, as the BH rotation increases, the amplitudes of the condensation profiles decrease. This suppression suggests that rotational effects act as an effective mechanism opposing the formation of the superconducting phase. Such behavior is consistent with previous holographic studies employing rotating BH backgrounds, though in asymptotically different spacetimes, which have likewise reported a reduction of the condensate with increasing rotation \cite{Lin:2014tza, Herrera-Mendoza:2024vfj, Srivastav:2019ixc}. Our results reinforce the robustness of this phenomenon across distinct holographic setups, highlighting the role of the background rotation in the condensate.
}


{
In addition to the condensation analysis, we now examine the electric conductivity \(\sigma_{\tiny{\mathrm{DC}}}\), which encodes the linear response of the dual system to small electric perturbations. In the holographic superconductor framework of Refs. \cite{Hartnoll:2008vx,Hartnoll:2008kx}, the optical conductivity is obtained by turning on time-dependent perturbations of the bulk gauge field and extracting the retarded Green’s function of the boundary current. In our setup, we consider fluctuations of the spatial components $A_{\theta}$ and $A_{\varphi}$, denoted by $\delta A_{\theta}$ and $\delta A_{\varphi}$, at the AdS boundary. Using the ans\"{a}tze (see Refs. \cite{Melnikov:2012tb, Hartnoll:2008vx})
\[
\delta A_{\theta,\varphi}=A_{\theta,\varphi}(r)e^{-i\omega t},\qquad \delta\Psi=0,
\tag{60}
\]
where $A_{\theta,\varphi}$ jointly denotes the $\theta$ and $\varphi$ components and $\omega$ is the frequency, and introducing the complex fluctuation $A = A_{\theta} d\theta + i A_{\varphi} d\varphi$, the Maxwell equations $\nabla_{\mu}F^{\mu\nu}=0$ reduce on the rotating axionic background (\ref{eq:metric}) to:
\[
A''_{\theta,\varphi}
+\frac{1}{2}\left(\frac{f'}{f}+\frac{h'}{h}\right)A'_{\theta,\varphi}
+\frac{\omega^2}{f\,h}A_{\theta,\varphi}=0,
\tag{61}
\]
with
\[
h(r)=f(r)N(r)^2-r^2\big(N^{\theta}(r)\big)^2.
\]
The function $h$ encodes the effect of the rotation function $N^{\theta}$, which is supported by the non-minimally coupled scalar field $\phi$ and the axionic field $\psi$. In contrast with the planar Schwarzschild-AdS$_4$ BH of Refs. \cite{Hartnoll:2008vx,Hartnoll:2008kx}, here the rotation and axion sector introduce a non-trivial mixing of temporal and spatial directions and provide an intrinsic mechanism for momentum dissipation in the boundary theory. Following the standard linear-response prescription  \cite{Hartnoll:2008vx,Hartnoll:2008kx}, an external oscillating electric field $E_i(\omega)$ (with $i=\{\theta,\varphi\}$) induces a boundary current $J_i(\omega)$, and the complex conductivity is defined as
\[
\sigma_{\tiny{\mbox{DC}}}(\omega)=\frac{J_{\theta,\varphi}(\omega)}{E_{\theta,\varphi}(\omega)}.
\]
Holographically, the electric conductivity is related to the retarded Green’s function of the boundary current via the Kubo formula \cite{Baggioli:2016rdj}:
\[
\sigma_{\tiny{\mbox{DC}}}(\omega)
=-\frac{i}{\omega}G^{R}_{J_{\theta,\varphi}J_{\theta,\varphi}}(\omega)
=\frac{A'_{\theta,\varphi}}{i\omega A_{\theta,\varphi}}\Big|_{r\to +\infty}.
\]
Here, the last equality follows from the standard near-boundary expansion of the bulk gauge field. The ingoing boundary condition at the horizon encodes dissipation in the dual field theory. In contrast, the asymptotic expansion at the AdS boundary determines the source (electric field) and the response (current).
\begin{figure}[h]
\centering
\includegraphics[width=0.48\textwidth]{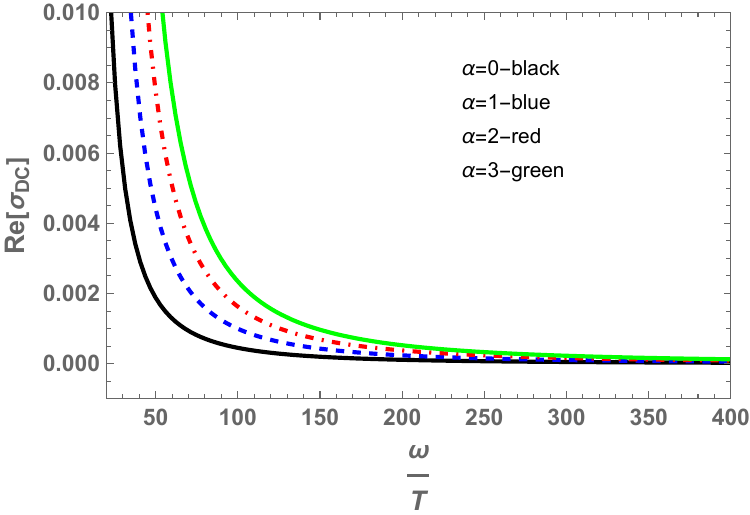}
\includegraphics[width=0.48\textwidth]{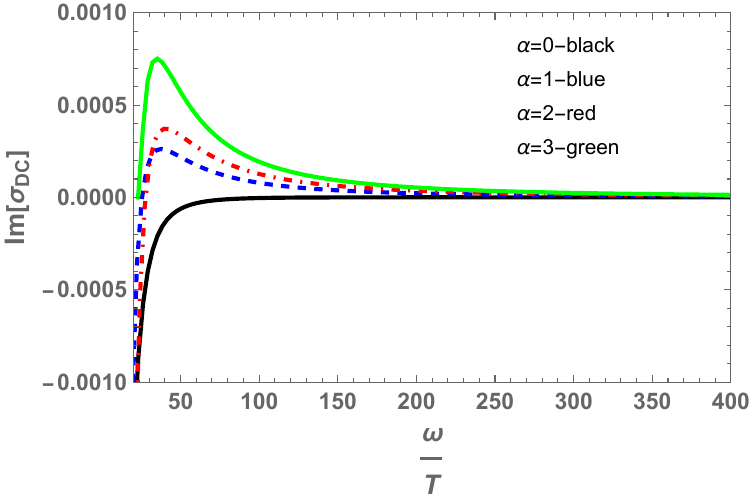}
\caption{The $\sigma_{\tiny{\mbox{DC}}}$ curve for the real (left panel) and imaginary (right panel) parts of the electrical conductivity for different values of the rotation parameter $\alpha$. The continuous (black) curve corresponds to the static case, while the other values correspond to situations with non-zero angular momentum.}
\label{fig:electrical}
\end{figure}

Figure \ref{fig:electrical} shows the real and imaginary parts of $\sigma_{\tiny{\mbox{DC}}}$ for different values of the rotation parameter $\alpha$. For $\alpha=0$ (the continuous black curve), the behavior is qualitatively similar to that of the original planar holographic superconductor (see Refs.  \cite{Hartnoll:2008vx,Hartnoll:2008kx,Gubser:2008px,Horowitz:2008bn}) where the real part develops a pronounced low-frequency enhancement, while the imaginary part displays a simple pole at $\omega=0$. For the situation $\alpha\neq 0$, the rotation deforms this pattern. The pole in $\mathrm{Im}(\sigma_{\tiny{\mbox{DC}}}(\omega))$ at $\omega=0$ persists, signalling a $\delta(\omega)$ contribution in $\mathrm{Re}(\sigma_{\tiny{\mbox{DC}}})$ through the Kramers–Kronig relations and thus the presence of a dissipationless component of the current, associated with the holographic superconducting phase, exactly as in Refs. \cite{Hartnoll:2008vx,Hartnoll:2008kx,Hartnoll:2009sz}. In the present model, however, the origin of this pole is not solely tied to momentum conservation. The axionic sector breaks translational invariance, leading to momentum relaxation in the dual theory and allowing charge carriers to dissipate their momentum into the effective “lattice’’ generated by the axion profile, in close analogy with linear-axion holographic models of momentum dissipation \cite{Andrade:2013gsa}. From a condensed-matter perspective, this mimics the effects of impurities, crystalline lattices, and disorder \cite{Amoretti:2019cef,Ammon:2019wci,Andrade:2017lsc}. The rotating axionic background thus provides a controlled holographic realization of a superconductor with both rotation and momentum dissipation. As the constant parameter $\alpha$ increases, the low-frequency structure of $\mathrm{Re}(\sigma_{\tiny{\mbox{DC}}})$ and the strength of the \(1/\omega\) pole in $\mathrm{Im}(\sigma_{\tiny{\mbox{DC}}})$ are modified, reflecting a non-trivial interplay between rotation, axion-induced momentum relaxation, and the superconducting condensate. This qualitative behavior is consistent with the suppression of the condensate by rotation observed in other holographic models based on rotating BHs \cite{Herrera-Mendoza:2024vfj,Lin:2014tza,Srivastav:2019ixc}, but arises here in a background with non-minimal scalar coupling and an explicit axionic momentum-dissipation mechanism.

Therefore, the numerical profiles in Fig. \ref{fig:electrical} show that the rotating axionic BH  provides a genuinely new setting for holographic superconductivity. The conductivity encodes how the rotation constant parameter $\alpha$, the axionic sector and the scalar couplings jointly determine the transport properties of the dual strongly coupled system, extending the original planar Schwarzschild–AdS$_4$ model of \cite{Hartnoll:2008vx,Hartnoll:2008kx,Gubser:2008px,Horowitz:2008bn} to a rotating, momentum-relaxing background, in the broader spirit of holographic transport and strange-metal phenomenology.
}


\section{Conclusions and discussions}\label{conclusions}

{In this work, we have constructed a new} axionically charged spinning {BH} in four dimensions with a non-minimally coupled scalar field. The above is possible via the inclusion of a suitable potential as well as a structural function coupling to the axionic field. {Although} the metric function $f$ and the rotating function $N^{\theta}$ enjoy logarithmic behavior, this configuration is asymptotically AdS. These configurations extend previous three-dimensional results and open avenues for further analysis of thermodynamics, stability, and holographic applications. 

As can be seen in Fig. \ref{fig1} (left panel), this configuration is characterized by two {constant} parameters ($\alpha$ and $\eta$) and only one {positive} integration constant $B$, allowing us to obtain one or even two horizons. Furthermore, we find that the metric function $f$ and the scalar field $\phi$ do not vanish for $\alpha \rightarrow 0^+$ and $\eta \neq 0$. This allows us to obtain a new static hairy BH configuration with one horizon (see Fig. \ref{fig1} (right panel)). 

{The presence of the scalar and axionic fields introduces additional
thermodynamic variables. Using the Euclidean action (\ref{euclideanaction}), we derived} the mass $\mathcal{M}$, Hawking temperature $T$, angular momentum $\mathcal{J}$, angular velocity $\Omega$, as well as the axionic potential {$\Psi_{a}$}, and axionic charge $\omega_a$, satisfying {the} first law (\ref{eq:first-law}). { It is straightforward to check that upon the following dimensional scaling $r_h \rightarrow \tilde{\lambda} r_h$ and $B \rightarrow \tilde{\lambda} B$, we have:
$${\cal{M}} \rightarrow \tilde{\lambda}^3 {\cal{M}},\quad {\cal{S}} \rightarrow \tilde{\lambda}^2 {\cal{S}},\quad {\cal{J}} \rightarrow \tilde{\lambda}^3 {\cal{J}},\quad \omega_{a} \rightarrow \tilde{\lambda} \omega_{a},$$ yielding the four-dimensional Smarr relation \cite{Smarr:1972kt}
\begin{eqnarray}\label{smarr}
\mathcal{M}=\frac{2}{3} T\mathcal{S}+\Omega \mathcal{J}+\frac{1}{3} \Psi_{a} \omega_{a}.
\end{eqnarray}}

In exploring the holographic superconductor description of these spinning BHs, we employed a numerical approach to investigate the condensation of the order parameter as a function of temperature and the rotation parameter (see Fig. \ref{fig:Condensate}). Notably, it is observed that the presence of a logarithmic falloff in the blackening function does not influence the {characteristic behavior} of the condensate. Nevertheless, our results reveal that the BH rotation impacts the order parameter. Specifically, we found that an {increase} in the {BH} rotation corresponds to a suppression in the amplitudes of the condensation curves, suggesting that the rotation is responsible for inhibiting the superconducting phase in our holographic system. This effect becomes relevant only at sufficiently high values of the rotation parameter. 

Together with the aforementioned, the electric conductivity as a function of the frequency $\omega$ is calculated numerically from the spinning configuration through the response of the system to small external sources. This is followed by the calculation of the Green's functions for the electric currents and the application of the Kubo formula. In this framework, the presence of the axionic field in breaking translational symmetry is crucial for modeling realistic condensed matter systems with momentum dissipation. In our case,  $\sigma_{\tiny{\mbox{DC}}}$ is influenced at the moment to consider the effects of rotation, represented through the parameter $\alpha$, as shown in Fig. \ref{fig:electrical}.

Some interesting directions for future research include, for example, the extension of spinning configurations following this toy model in higher dimensions and exploring a broader range of values for the non-minimal coupling constant $\xi$.  {Axion BHs solutions have also been constructed in the context of effective string theory, most notably in the works of
Gal'tsov, Kechkin et al. ~\cite{Galtsov:1994pd,Galtsov:1995zu,Garcia:1995qz}. Although these configurations allow us valuable insights into the role of axionic fields within gravitational backgrounds, their construction is based on the absence of a cosmological constant and relies on specific solution-generating symmetries that do not directly apply to the AdS case considered in the present work. Establishing corresponding solutions in the AdS context would necessitate a separate analysis. Nonetheless, it would be intriguing to explore these generalizations in future research.} 
Additionally, we can obtain a modification of the first law of BH thermodynamics, considering now the cosmological constant  $\Lambda$ as a thermodynamic parameter related to the pressure $P$ of the system \cite{Kastor:2009wy,Kubiznak:2014zwa,Kubiznak:2016qmn,Dolan:2010ha,Cvetic:2010jb}, where now the first law takes the form
$${\delta {\cal{M}}=T \delta {\cal{S}}+V\delta P+\Omega \delta {\cal{J}}+\Psi_a \delta {\omega_a}}.$$
Here, $V$ corresponds to the conjugate thermodynamic
volume, and $\cal{M}$ now takes the role of the enthalpy. \textcolor{black}{Along with the above, an interesting open problem is to explore this when $\Lambda$ is included as one of the dynamical variables within the AdS/CFT framework, where, from a holographic point of view, the Smarr formula (\ref{smarr}) can be extended and understood (see Refs. \cite{Mancilla:2024spp,Karch:2015rpa}).}

\begin{acknowledgements}
{The authors would like to thank the anonymous referees for carefully reading our manuscript and giving valuable suggestions that led to an improved version of this work.} The authors gratefully acknowledge the insightful comments of Professor Julio Oliva. 
M.B. is supported by proyecto interno UCM-IN-25202. JAHM and DFHB are also grateful to SNII from CONAHCYT. 
\end{acknowledgements}
\section*{Conflict of Interest}
The authors declare no conflict of interest. 
\section*{Data Availability Statement}
Data sharing is not applicable to this article, as no data sets were generated or analyzed during the current study.

\appendix

\section{Equations of motion}\label{EM}

In this section, we present the explicit expression for the equations of motion given in (\ref{eq:field_eqs}) together with the line element (\ref{eq:metric}) for $\xi=\frac{1}{6}$ and $N(r)=1$. The non-vanishing components read:
\begin{eqnarray*}
\mathcal{E}_{\phantom{a} t}^{t}&=&2 r N^{\theta}(N^{\theta})'-\frac{1}{3} r N^{\theta}(N^{\theta})'\phi^2+\frac{1}{2}r^2 N^{\theta} (N^{\theta})''-\frac{1}{12}r^2 N^{\theta}(N^{\theta})''\phi^2-\frac{1}{6} r^2 \phi \phi' N^{\theta} (N^{\theta})'\\
&+&\frac{f}{r^2}- \frac{f \phi^2}{6 r^2}+\frac{f'}{r}-\frac{f' \phi^2}{r}
+\frac{1}{4}r^2\big((N^{\theta})'\big)^2-\frac{1}{24} r^2 \big((N^{\theta})'\big)^2 \phi^2 -\frac{1}{6} \phi \phi' f'+\frac{1}{6}(\phi')^2f\\
&-&\frac{1}{3} f \phi \phi''-\frac{2}{3 r} \phi \phi' f+U(\phi)+\Lambda,
\end{eqnarray*}
\begin{eqnarray*}
\mathcal{E}^{t}_{\phantom{a} \theta}&=&2 (N^{\theta})' r-\frac{1}{3} r (N^{\theta})'\phi^2+\frac{1}{2}r^2(N^{\theta})''-\frac{1}{12} r^2 (N^{\theta})''\phi^2-\frac{1}{6}\phi \phi' r^2((N^{\theta}))',\\ \\
\mathcal{E}^{\theta}_{\phantom{a} t}&=&-\frac{1}{2} (N^{\theta})'' f-\frac{1}{2} r^2 (N^{\theta})'(N^{\theta})^2+\frac{1}{12} (N^{\theta})'' f \phi^2+\frac{1}{12} r^2 (N^{\theta})''\phi^2 (N^{\theta})^2+\frac{1}{2} N^{\theta} f''\nonumber\\
&-&\frac{1}{12} N^{\theta} f'' \phi^2-r^2 N^{\theta}\big((N^{\theta})'\big)^2+\frac{1}{6} r^2 N^{\theta} \big((N^{\theta})'\big)^2\phi^2+\frac{1}{3r} (N^{\theta})' f \phi^2+\frac{1}{3}r(N^{\theta})'\phi^2 (N^{\theta})^2\\
&-&\frac{2 (N^{\theta})' f}{r}-2 r(N^{\theta})^2 (N^{\theta})'+\frac{1}{6} (N^{\theta})' \phi' \phi f+\frac{1}{6} r^2 (N^{\theta})' \phi' \phi (N^{\theta})^2+\frac{1}{3r} N^{\theta}\phi' \phi f\\
&-&\frac{1}{6}N^{\theta} \phi' \phi f'-\frac{N^{\theta} f}{r^2}+\frac{1}{6r^2} N^{\theta} f \phi^2,
\end{eqnarray*}
\begin{eqnarray*}
\mathcal{E}^{r}_{\phantom{a} r}&=&-\frac{1}{2}(\phi')^2 f-\frac{2}{3r}\phi \phi' f-\frac{1}{6}\phi \phi' f'-\frac{1}{6r} f'\phi^2+\frac{f'}{r}+\frac{1}{4} r^2 \big((N^{\theta})'\big)^2-\frac{1}{24} r^2 \big((N^{\theta})'\big)^2\phi^2\nonumber\\
&-&\frac{1}{6r^2} f \phi^2+\frac{f}{r^2}+U(\phi)+\Lambda,
\end{eqnarray*}

\begin{eqnarray*}
\mathcal{E}^{\theta}_{\phantom{a} \theta}&=&\frac{1}{2} r^2 N^{\theta} (N^{\theta})''+\frac{1}{12} r^2 N^{\theta} (N^{\theta})''\phi^2+\frac{1}{2} f''-\frac{1}{12} f''\phi^2-\frac{1}{3} f \phi \phi''-\frac{3}{4} r^2\big((N^{\theta})'\big)^2\\
&+&\frac{1}{8} r^2 \big((N^{\theta})'\big)^2\phi^2+\frac{1}{3} r N^{\theta} (N^{\theta})' \phi^2
-2 r N^{\theta} (N^{\theta})'+\frac{1}{6} r^2 N^{\theta}((N^{\theta})')\phi' \phi+\frac{1}{6}(\phi')^2 f \\
&-&\frac{1}{3r}\phi \phi' f-\frac{1}{3} \phi \phi' f'+\frac{f'}{r}-\frac{1}{6r} f' \phi^2+U(\phi)+\Lambda,\\ \\
\mathcal{E}_{\phantom{a} \varphi}^{\varphi}&=&\frac{1}{2} f'' -\frac{1}{12} f'' \phi^2-\frac{1}{3} f \phi \phi''+\frac{1}{6} (\phi')^2 f-\frac{1}{3r} \phi \phi'f-\frac{1}{3}\phi \phi'f'+\frac{f'}{r}-\frac{1}{6r} f' \phi^2\nonumber\\
&-&\frac{1}{4} r^2 \big((N^{\theta})'\big)^2+\frac{1}{24} r^2 \big((N^{\theta})'\big)^2\phi^2+U(\phi)+\Lambda,
\end{eqnarray*}
where, as before,  ($')$ denotes the derivative with respect to coordinate $r$. On the other hand, the explicit expression for the Einstein equations $\mathcal{G}_{\mu \nu}:= \mathcal{E}_{\mu\nu}- T^{\psi}_{\mu\nu}$, given in eq. (\ref{eq:field_eqs-final}), follow directly from the above expressions, together with $T^{\psi\,t}_{\phantom{\mu \eta}t}=T^{\psi\,r}_{\phantom{\mu \eta}r}=T^{\psi\,\theta}_{\phantom{\mu \eta}\theta}=-T^{\psi\,\varphi}_{\phantom{\mu \eta}\varphi}=-\frac{1}{2 r^2}\varepsilon (\partial_{\varphi} \psi)^2,$ where, {as before} $\partial_{\varphi} \psi:=\partial \psi/\partial \varphi.$

\section{Conserved charges $\cal{M}$ and $\cal{J}$}\label{termo-b}

{To compute the conserved charges associated with the solution, namely the mass and angular momentum, we employ the Brown–York quasilocal formalism, where a key element is the stress tensor. Starting from the four-dimensional line element (\ref{eq:metric}) with $N=1$, we consider the surface
at a fixed radius $r=R=\mbox{constant}$ as the boundary $\partial {\cal M}$ with the induced metric $\gamma_{a b}$. The renormalized action $I_{\tiny{\mbox{ren}}}$ with a negative cosmological constant (\ref{eq:Lambda}) takes the form:
$$I_{\tiny{\mbox{ren}}}=I_{\tiny{\mbox{bulk}}}+\int_{\partial_{\cal{M}}} d^3 x \sqrt{-\gamma} \left\{\left(1-\frac{1}{6} \phi^2\right)K+\frac{2}{\ell}\left(1-\frac{1}{6} \phi^2\right)-\frac{\ell}{2} \left(1-\frac{1}{6} \phi^2\right){\cal R}[\gamma]\right\}.$$
Here,  $I_{\tiny{\mbox{bulk}}}$ is the bulk action, $\gamma$ is the determinant of the induced metric at the boundary, 
$K=\gamma^{a b} K_{a b}$ is the trace of the extrinsic curvature $K_{a b}$. Here, $n_{a}=\frac{dr}{\sqrt{f}}\Big{|}_{r=R}$, is the outward-pointing unit normal vector to $\partial {\cal M}$, ${\cal R}[\gamma]$ is the scalar curvature of the boundary metric and $R[\gamma]=0$ for the planar boundary considered here.}

{The stress tensor is defined as $T_{a b}=({2}/{\sqrt{-\gamma}}) ({\delta I_{{\tiny{\mbox{ren}}}}}/{\delta \gamma^{a b}})$ and takes the form: $$T_{a b}=\left(1-\frac{1}{6} \phi^2\right) (K_{a b}-K \gamma_{a b})-\left(\frac{2}{\ell}\right)\left(1-\frac{1}{6} \phi^2\right)\gamma_{a b}+T_{a b}^{{\tiny{\mbox{mat}}}},$$
where $T_{a b}^{\tiny{\mbox{mat}}}$ contains contributions from matter counterterms. In the present solution, these additional terms do not contribute to the conserved charges, since the scalar field vanishes asymptotically and the axionic sector only affects the charges through its backreaction on the metric functions.}

{On a spacelike surface $\Sigma$ in $\partial {\cal M}$ with metric $h_{ij}$, the induced metric $\gamma_{a b}$ is written according to the  Arnowitt, Deser, Misner (ADM) decomposition in the following form:
\begin{eqnarray*}
\gamma_{ab} dx^a dx^b &=&-N^2_B dt^2+h_{ij} (dx^i+N^i dt)(dx^j+N^j dt), \qquad i,j \in\{\theta, \varphi\},\\
N^2_B&=&f-r^2(N^{\theta})^2 \big{|}_{r=R},
\end{eqnarray*}
where $N_B$ and $N^i=(N^{\theta}{|}_{r=R},0)$ are the lapse function and shift vector, respectively. The  timelike unit vector $u^a$ normal to $\Sigma$ reads: 
$$u^a=\frac{1}{N_B} \left((\partial_t)^a-N^{\theta}(\partial_\theta)^a\right).$$
If $\xi^{a}$ is a Killing vector generating an isometry of the boundary geometry, there exists an associated conserved charge. Following the Brown-York method \cite{Brown:1992br,Brown:1994gs,Balasubramanian:1999re}, this charge is given by:
\begin{equation*}
Q_{\xi}=\int_{\Sigma} d^2 x \sqrt{h}\, u^{a} T_{a b} \xi^{b}.
\end{equation*}
Here, taking a rotational Killing vector $\partial_\theta = -\xi^{a} \partial_{a}$ we obtain:
$$J(R)=\int_{\Sigma} d^2 x \sqrt{h}\, u^{a} T_{a \theta} \simeq -\frac{\sigma \sqrt{\alpha }  B^3}{12}+O\left(\frac{1}{R^2}\right),$$
and the angular momentum can be obtained as
$${\cal J}=\lim_{R\rightarrow +\infty} J(R)=-\frac{\sigma}{12} {\sqrt{\alpha}B^3},$$
recovering the expression given in eq. (\ref{momentum-vel}). The nonvanishing value of this conserved charge provides an explicit proof that the configuration carries angular momentum and therefore corresponds to a genuinely rotating BH solution. Similarly, with a timelike Killing vector $\partial_t = \xi^{a} \partial_{a}$:
$$M(R)=\int_{\Sigma} d^2 x \sqrt{h}\, u^{a} T_{a t} \simeq \frac{\sigma\eta B^3}{18}+O\left(\frac{1}{R^2}\right),$$
and the mass $\cal{M}$ takes the structure
$${\cal M}=\lim_{R\rightarrow +\infty} M(R)=\frac{\sigma \eta B^3}{18},$$
being consistent with eq. (\ref{mass}).
}


\end{document}